\documentclass[%
 reprint,
 amsmath,amssymb,
 aps,
twocolumn]{revtex4}

\usepackage{graphicx}
\usepackage{dcolumn}
\usepackage{bm}
\usepackage{color}
\usepackage{float}

\def\sideremark#1{\ifvmode\leavevmode\fi\vadjust{\vbox to0pt{\vss
 \hbox to 0pt{\hskip\hsize\hskip1em
 \vbox{\hsize2cm\tiny\raggedright\pretolerance10000
  \noindent #1\hfill}\hss}\vbox to8pt{\vfil}\vss}}}

\newcommand{\R}{\mathbb R}
\newcommand{\C}{\mathbb C}
\newcommand{\cg}{\mathfrak g}
\newcommand{\tens}{\otimes}
\newcommand{\extd}{{\rm d}}

\newcommand{\eps}{\varepsilon}

\newcommand{\del}{\partial}
\newcommand{\Tr}{{\rm Tr}}
 \newcommand{\<}{\langle}
  \renewcommand{\>}{\rangle}
  \newcommand{\CL}{\mathcal L}

\begin{document}

\title{Kaluza-Klein ansatz from Lorentzian quantum gravity on the fuzzy sphere}


\author{Chengcheng Liu and Shahn Majid}
\affiliation{
School of Mathematical Sciences\\
Queen Mary University of London\\  Mile End Rd, London E1 4NS, UK}

\email{chengcheng.liu@qmul.ac.uk, s.majid@qmul.ac.uk} 
\thanks{CL was  supported by a China Scholarship and SM by Leverhulme Trust project grant RPG-2024-177}

\date{\today}

\begin{abstract} If Kaluza-Klein ideas were correct as an explanation of Yang-Mills and General Relativity on spacetime, the extra fibre geometry would have to be a sphere of constant size of the order of 10 Planck lengths, hence subject to quantum gravity corrections. Conversely, it was shown in previous work that modelling such corrections by noncommutative coordinates indeed forces the Kaluza-Klein cylinder ansatz form of the metric, and we now propose that the remaining restrictions needed come from quantum gravity on the fibre. Working with a fuzzy sphere fibre, we find that the expected value of the metric is indeed spherical and we propose that it can be taken as of constant size due to freedom in the renormalisation of divergences. In this way, we outline a mechanism whereby the observed structure of gravity plus Yang-Mills  can emerge at low energies as a consequence of quantum gravity effects. \end{abstract}
\maketitle 

\section{Introduction}\label{secintro}

The `Kaluza-Klein miracle' as it first emerged was a stunning idea whereby electromagnetism, i.e. the Maxwell action, can be seen as part of the Einstein-Hilbert action on an extension of spacetime by an internal $S^1$ fibre at each point. The same can be done for Yang-Mills using as fibre a nonAbelian group or homogeneous space such as a 2-sphere at each point, see \cite{Coq}. Unfortunately, however,  the idea depends on a specific ansatz for the total space metric and other assumptions, so it is not so much that gravity on the extended space decomposes into gravity and gauge theory but just that it is merely a big enough canvas to contain those as special modes. By itself, this therefore lacks explanatory power. 

Moreover,  in order to match up with the observed coupling constant/vacuum permitivity for electro-magnetism, the radius of the $S^1$ in the original model has to be 23 Planck lengths, while for $SU(2)$ Yang-Mills, a similar calculation for the weak force requires 11 Planck lengths for the radius of a sphere at each point, see  \cite{LiuMa1,LiuMa2} for a recent recalculation. Being of Planckian order, one can expect that the fibre geometry is therefore subject to quantum gravity corrections. What these look like is not clear, but a currently plausible hypothesis is that these can be modelled by using `quantum' or noncommutative coordinates\cite{Sny, Ma:pla, MaRue, AmeMa, DFR, Hoo,BegMa}. This motivated \cite{ArgMa3,LiuMa1,LiuMa2,LiuMa3} to reexamine the Kaluza-Klein idea with a noncommutative geometry as fibre. This reasoning as well as our methodology using quantum Riemannian geometry(QRG) \cite{BegMa} is rather different from Connes approach to the Standard Model\cite{ChaCon} also using  noncommutative geometry, but not incompatible. 

What was found, particularly in \cite{LiuMa1,LiuMa2} using $M_2(\C)$ and the fuzzy sphere respectively for the fibre coordinate algebra, was that noncommutativity of the fibre exactly forces the `cylinder-ansatz' form of the metric whereby all its components are constant on the fibre. The fibre metric can still vary on spacetime and amounts to a matrix-valued Liouville-type  field. The precise statement is then that gravity on the total space exactly decomposes from the point of view of spacetime as a Yang-Mills-like field, gravity and this Liouville-type field. This goes a long way towards deriving or explaining the form of what we observe at low energies as being due to quantum gravity corrections. To complete this argument, however, we still need to know why the Liouville field should have value corresponding to the metric of a sphere of constant size, as needed to obtain exactly Yang-Mills with a spacetime-independent coupling constant. Such a gap is also present in the classical Kaluza-Klein argument where it is known that this not only cannot be derived from, but is incompatible with, the equations of motion for the Liouville field. In the present work, we propose that it instead emerges in an effective theory where we quantise the fibre geometry, which we illustrate in detail for the case of the fuzzy sphere. Quantum gravity on the fuzzy sphere was already studied in \cite{LirMa,Ma:qg} for Euclidean signature, but we will need the significantly harder Lorentzian version where there is an $\imath$ in the action. We show that this again has  divergences but that otherwise the expected value of the fibre metric is indeed  a sphere, and we then argue that the freedom from regularising and renormalising the divergences allows one to take the fuzzy sphere size at a given energy scale as constant. 

In Section~\ref{secpre} we provide a brief recap of quantum Riemannian geometry and how it leads to the cylinder ansatz in \cite{LiuMa2},  and  in Section~\ref{seclc} how this appears on spacetime as Yang-Mills, gravity and a matrix-Liouville field. In Section~\ref{seceqm}, we show that there is still the same inconsistency with the equations of motion of the latter as when taking a compact classical fibre, i.e. that this field cannot be treated classically if we aim to derive what we observe. Section~\ref{secqg} then treats it at each spacetime point in isolation as a Lorentzian quantum gravity theory on the fibre, with conclusions as discussed. Section~\ref{secmatch} then looks at matching the quantum gravity expectations to the required numerical values. While some aspects, especially the latter due to numerical noise, are not conclusive, this provides a scenario where the essential nature of the standard model as Yang-Mills/Maxwell plus gravity could arise from quantum gravity effects. 

\section{QRG origin of the cylinder ansatz} \label{secpre}

Without going into details, a modern way of working with a noncommutative `coordinate algebra' $A$ (and common to most approaches to noncommutative geometry\cite{Con,BegMa}) is to work with an exterior algebra  $\Omega(A)=\oplus_i\Omega^i$  of `differential forms', where $
\Omega^0=A$. We take $\extd:\Omega^i\to \Omega^{i+1}$ with its usual properties $\extd^2=0$ and $\extd$ a graded-derivation,  but without assuming that anything graded commutes. We also work over $\C$ with the `real geometry' encoded in a graded $*$-algebra structure on $\Omega(A)$. In this context, a quantum metric is $\cg\in \Omega^1\tens_A\Omega^1$  with certain properties, the most important for us being an `inverse metric'  $(\ ,\ ):\Omega^1\tens_A\Omega^1\to  A$ which is required to be a bimodule map and defined over $\tens_A$, i.e. 
\[ a(\omega, \eta)=(a\omega,\eta),\quad (\omega a, \eta)=(\omega, a\eta),\quad (\omega,\eta a)=(\omega,\eta)a\]
for all $a\in A$ and $\omega,\eta\in \Omega^1$. This is just a statement of tensoriality,  the same as in GR where in tensor calculus one can take functions past any tensor indices.  The critical thing for our argument is that these reasonable properties force that $\cg$ commutes with $a\in A$, $\cg a=a\cg$, even though in general 1-forms are not assumed to commute with functions\cite{BegMa}. We also assume that $\cg$ is quantum symmetric in a suitable sense.

We apply this to $A=C^\infty(M)\tens A_f$ and the (graded) tensor product exterior algebra, where $A_f$ is the `fibre' coordinate algebra and allowed to be noncommutative. We further assume that $\Omega^1(A_f)$ has a basis $\{s^i\}$ over $A_f$. Then the space of 1-forms on the product is $\Omega^1(M)\tens A_f\oplus C^\infty(M)\tens \Omega^1(A_f)$ and hence a general element of the tensor square has the local form\cite{ArgMa3,LiuMa1,LiuMa2}
\[ \cg=\extd x^\mu g_{\mu\nu}\tens \extd x^\nu + \extd x^\mu A_{\mu i}\tens s^i+s^i B_{\mu i}\tens\extd x^\mu+ s^i h_{ij}\tens s^j\]
for some coefficient fields in the product algebra. We require the metric to be suitably quantum symmetric  and in the present context this amounts to
\begin{equation}\label{prodsym} g_{\mu\nu}=g_{\nu\mu},\quad A_{\mu i}=B_{\mu i}\end{equation}  and an appropriate quantum symmetry for $A_f$. 

While not the most general case, the simplest nontrivial case is to assume that the $\{s^i\}$ are central (they commute with $a\in A_f$) and  that $A_f$ is `highly noncommutative' in the sense of trivial centre $Z(A_f)=\C 1$. Given that $\cg$ has to central, this implies that $g_{\mu\nu}, A_{\mu i}, h_{ij}$ depend only on spacetime  (they are proportional to $1\in A_f$), {\em which is exactly the Kaluza-Klein cylinder ansatz}.

\medskip 
\section{Decomposition of gravity on the product}\label{seclc}

Again without going into details, a gravitational connection is a map $\nabla:\Omega^1\to \Omega^1\tens_A\Omega^1$ with certain properties. We will want one that is torsion free and metric compatible i.e.  `quantum Levi-Civita'  in a natural sense\cite{BegMa}. Note that evaluating the left-most factor with a `quantum vector field' $\Omega^1\to A$ will result in a `covariant derivative' along  more more familiar lines. One then has a Riemannian curvature tensor and, subordinate to a splitting $i:\Omega^2\to \Omega^1\tens_A\Omega^1$ that expresses a 2-form as a tensor, one has a Ricci tensor as a trace of the Riemann tensor, and from this a Ricci scalar. The standard choice of $i$ if there is an anticommuting generating set of 1-forms (which will be our case) is to lift a 2-form to an antisymmetric product of 1-forms.  

Proceeding in the context of our product geometry, we let
\begin{align*} \tilde g^{\mu\nu}&=(\extd x^\mu,\extd x^\nu),\quad    \tilde h^{ij}=(s^i,s^j),\\
 \tilde A^{\mu i}&=-(\extd x^\mu,s^i)=\tilde g^{\mu\nu}h^{ij}A_{\nu j},\end{align*}
 with inverses 
 \[ \tilde g_{\mu\nu}=g_{\mu\nu}-h^{ij}A_{\mu i}A_{\nu j},\quad 
 \tilde h_{ij}=h_{ij}- g^{\mu\nu}A_{\mu i}A_{\nu j}.\]
Here $\tilde g,\tilde A$ will be the more relevant physical fields on spacetime $M$. To exhibit the gravitational connection and curvature, we specialise now to $A_f$ the fuzzy sphere case in \cite{LiuMa2}, but the ideas are more general. The algebra $A_f$  has self-adjoint generators which we denote $y^i$ with relations
\[[y^i,y^j]=2\imath\lambda \eps_{ijk}y^k,\quad \sum_i (y^i)^2=1-\lambda^2,\]
where $\lambda\ne 0$ is a real deformation parameter, $i,j,k$ run $1,2,3$, sum over $k$ is understood and $\eps_{ijk}$ is the totally antisymmetric tensor. The $n\times n$-matrix algebra fuzzy spheres in \cite{Str,Mad} can be seen as quotients of this for discrete values $\lambda=1/n$. The natural $\Omega^1(A_f)$ is
\[ \extd y^i= \eps_{ijk} y^j s^k,\quad [y^i,s^j]=0,\quad \extd s^i=-{1\over 2}\eps_{ijk}s^j\wedge s^k\]
with the $\{s^i\}$ forming a Grassmann algebra. Then under some mild technical assumptions, one has  a unique quantum Levi-Civita connection on the product\cite{LiuMa2}
\begin{widetext}
\begin{align*}
\nabla \extd x^\mu&=-\Gamma^\mu_{\alpha\beta}\extd x^\alpha\tens \extd x^\beta+F^\mu_{\alpha i}(\extd x^\alpha\tens s^i+s^i\tens \extd x^\alpha)+D^\mu_{ij} s^i\tens s^j,\\
\nabla s^k&=E^k_{\alpha\beta}\extd x^\alpha\tens \extd x^\beta+B^k_{\alpha i}(\extd x^\alpha\tens s^i+ s^i\tens \extd x^\alpha)+H^k_{ij} s^i\tens s^j\\
\Gamma^\sigma_{\mu\nu}&=\tilde\Gamma^\sigma_{\mu\nu}+\frac{1}{2}\tilde g^{\sigma\rho}(A_{\mu i} \del_{[\nu} (A_{\rho]j}h^{ij})+A_{\nu i} \del_{[\mu} (A_{\rho]j}h^{ij})+A_{i\mu}A_{j\nu}\del_\rho h^{ij}),\end{align*}
\end{widetext}
where
\begin{align*}
F^\mu_{\alpha i}&=\frac{1}{2}\tilde g^{\beta\mu}(-\del_{[\alpha}A_{\beta]i}+h^{jk}A_{k\beta}(\del_\alpha h_{ij}-A_{\alpha l}\epsilon_{ijl})),\\
D^\mu_{ij}&=\frac{1}{2}\tilde g^{\mu\alpha}(\del_\alpha h_{ij}-A_{k\alpha}h^{mk}h_{n(i}\epsilon_{j)mn}), \\
E_{\alpha \beta}^k&=-\frac{1}{2}h^{ki}\nabla_{(\alpha} A_{ \beta)i}, \\
B^k_{\alpha i}&=\frac{1}{2}\tilde h^{jk}(g^{\mu\beta}A_{j\mu}\del_{[\alpha}A_{\beta]i}+A_{l\alpha}\epsilon_{ijl}-\del_\alpha h_{ij}),\\
H^l_{ik}&=\frac{1}{2}\tilde h^{jl}(h_{n(i}\epsilon_{k)jn}-A_{j\mu}g^{\mu\alpha}\del_\alpha h_{ik})-\frac{1}{2}\epsilon_{lik}
\end{align*}
and $\tilde\Gamma^\sigma_{\mu\nu}$ are the classical Levi-Civita connection coefficients for the effective metric $\tilde g$. Here $\nabla_\alpha$ is the covariant derivative defined by $\Gamma$ given above. The  Ricci scalar on the product then comes out as\cite{LiuMa2}
\begin{align*}
\ R=&\tilde R_M+R_h+\frac{1}{8}h_{ij}\tilde F^i_{\mu\nu}\tilde F^{j\mu\nu}\\
&+\frac{1}{2}\tilde\nabla^\alpha {\rm Tr}(\Phi_\alpha)+\frac{1}{8}\big({\rm Tr}(\Phi_\alpha\Phi^\alpha)+{\rm Tr}(\Phi_\alpha){\rm Tr}(\Phi^\alpha)\big)
\end{align*}
where
\begin{align*}\Phi^i_{\alpha j}&:=h^{ik}\tilde\nabla_{A\alpha} h_{kj},\  \Phi^{\alpha i}_{\ \ j}:=h^{ik}\tilde\nabla^\alpha_{\ A} h_{kj},\  \tilde\nabla^\alpha_{\  A}:=\tilde g^{\alpha\beta}\tilde\nabla_{\beta  A}\\ 
\tilde\nabla_{\alpha A}&(f_{i_1 i_2...}):=\tilde\nabla_{\alpha}(f_{i_1i_2i_3...})\\
&\ -\tilde A_{\alpha j}(\epsilon_{i_1jk}f_{ki_2i_3...}+\epsilon_{i_2jk}f_{i_1ki_3...}+\epsilon_{i_3jk}f_{i_1i_2k...}+...)\end{align*}
for any $f_{i_1 i_2 i_3...}$ with latin indices. $\tilde R_M$ is the  Ricci scalar on $M$ for the physical metric $\tilde g$ and
\[R_{h}={e^{-\Tr(\Phi)}\over 2}\Big(\Tr(e^{2\Phi})-{1\over 2}\Tr(e^\Phi)^2\Big) \]
is the Ricci scalar on the fuzzy sphere regarded as a potential term for 
\[ \Phi=\ln(\underline h)\]
as a matrix $\underline h=\{h_{ij}\}$, while $\tilde F$ is the field strength of $\tilde A$:
\begin{equation}
\tilde F^{i\mu\nu}=\del^{[\mu}\tilde A^{\nu]i}-\tilde A^{\mu j}\tilde A^{\nu k}\epsilon^{ijk}, \quad \tilde F^i_{\mu\nu}=\tilde g_{\alpha\mu}\tilde g_{\beta \nu} \tilde F^{i\alpha\beta}.
\end{equation}

There is also a natural integration $\int_{A_f}$ given by averaging over all rotations of an element (the quantum geometry was designed to retain a classical $SU(2)$ invariance). Then the Einstein-Hilbert action in the product amounts to
\begin{align}\label{action}
S&=\int_M\extd^n x V_f \sqrt{-|\tilde g|}\Big(\tilde R_M+R_h+\frac{1}{8}h_{ij}\tilde F^i_{\mu\nu}\tilde F^{j\mu\nu}\nonumber\\
&\quad+\frac{1}{2}\tilde\nabla^\alpha {\rm Tr}(\Phi_\alpha)+\frac{1}{8}\big({\rm Tr}(\Phi_\alpha\Phi^\alpha)+{\rm Tr}(\Phi_\alpha){\rm Tr}(\Phi^\alpha)\big)\Big)
\end{align}
viewed as an action for fields on spacetime $M$, where $V_f=\int_{A_f}1$ is the volume of the fuzzy sphere. This is because the Ricci scalar on the product is independent of the fuzzy sphere. Here  $h_{ij}$ is a kind of matrix-valued `Liouville-sigma model' field in view of the form of its potential. Similar results can be obtained for a classical $S^3$ fibre\cite{Coq}, even though the derivation is quite different. Now, however, this is the exact content of gravity on the product, rather than the content of an ansatz.

\section{Problem with the equations of motion}\label{seceqm}

We still need to have that $h_{ij}=h\delta_{ij}$ for a constant $h$ if we are to recover regular Yang-Mills and gravity with constant couplings (i.e. not varying on spacetime). As for Kaluza-Klein theory with classical compact fibre, the equations of motion for the above action (\ref{action}) do not, unfortunately, lead to this. They can be computed as
\begin{align}
&\tilde R_{\alpha\beta}-\frac{1}{2}\tilde R_M \tilde g_{\alpha\beta}=-\frac{1}{2}(T^{YM}_{\alpha\beta}+T^\Phi_{\alpha\beta})\\
&\tilde \nabla^\alpha_{\  A}(V_f h_{ij}\tilde F^{j}_{\alpha\beta})=\epsilon_{ijk}\Phi^j_{\beta k}V_f\\
&\frac{1}{V_f}\tilde \nabla^\alpha_{\ A}(V_f(\Phi^i_{\alpha j}+\delta^i_j {\rm Tr}(\Phi_\alpha)))\nonumber\\
&=\frac{1}{2}h_{jk}\tilde F^k_{\mu\nu}\tilde F^{i\mu\nu}+2 h^{ik}T^h_{kj}-4h^{ik}\frac{\del \ln V_f}{\del h^{jk}}R+\frac{2\delta^i_{j}}{V_f}\tilde\square_A V_f\label{EL}
\end{align}
where
\begin{align}
T^h_{ij}&=\delta_{ij}+{1\over 2}h^{lk}h_{mn}\epsilon_{jmk}\epsilon_{iln}+\frac{1}{4}h_{ia}h_{jb}h^{mn}h^{lk}\epsilon_{bmk}\epsilon_{anl}\\
T^{YM}_{\alpha\beta}&=\frac{1}{2}\left(\tilde g^{\mu\nu}{\rm Tr}(\tilde F_{\alpha \mu}e^\Phi \tilde F_{\beta \nu})-\frac{1}{4}\tilde g_{\alpha\beta}{\rm Tr}(\tilde F_{\mu\nu}e^\Phi \tilde F^{\mu\nu})\right)\\
T^\Phi_{\alpha\beta}&=- R_h \tilde g_{\alpha\beta}-\frac{1}{8}\tilde g_{\alpha\beta}({\rm Tr}(\Phi_\mu\Phi^\mu)+{\rm Tr}(\Phi_\mu) {\rm Tr}(\Phi^\mu))\nonumber\\
&\quad +\frac{1}{4}({\rm Tr}(\Phi_\alpha\Phi_\beta)+{\rm Tr}(\Phi_\alpha) {\rm Tr}(\Phi_\beta))\nonumber\\
&\quad +\frac{1}{2} (\tilde g_{\alpha\beta}{\rm Tr}(\Phi_\mu)\del^\mu\ln V_f-{\rm Tr}(\Phi_{(\alpha})\del_{\beta)}\ln V_f).
\end{align}
Here, the total energy-momentum tensor $T^{YM}_{\alpha\beta}+T^\Phi_{\alpha\beta}$ is conserved, while $T^h{}_{ij}=-2{\delta R_h\over\delta h^{ij}}$ (albeit, currently without a  theory of noncommutative variational calculus). But if we set  $h_{ij}=h\delta_{ij}$ with $h$ constant then (\ref{EL}) becomes 
\[ \tilde F^i_{\mu\nu} \tilde F^{j\mu\nu}=-{2\over h^2}\delta^{ij}\]
so that the YM curvature is unreasonably constrained by the equations of motion.  In line with the more usual Kaluza-Klein literature\cite{KK}, one could also suppose at outset that $h_{ij}=h\delta_{ij}$ with $\phi=\ln(h)$ is a single Liouville field. In this case, the action and the ensuing equations of motion simplify to \begin{align*}
S&=\int_M\extd^n x V_f\sqrt{-|\tilde g|}\\
&\kern-20pt \left(\tilde R_M+\frac{1}{8}e^\phi\tilde F^i_{\mu\nu}\tilde F^{i\mu\nu}+\frac{3}{2}(\tilde \nabla^\alpha\tilde \nabla_\alpha\phi+\del_\alpha \phi\del^\alpha \phi-\frac{1}{2}e^{-\phi})\right)\\
\tilde R_{\alpha\beta}&-\frac{1}{2}\tilde R_M \tilde g_{\alpha\beta}=-\frac{1}{2}(e^\phi T^{YM}_{\alpha\beta}+T^\phi_{\alpha\beta})\\
\tilde \nabla^\alpha_{\  A}&(V_f e^\phi\tilde F^{i}_{\alpha\beta})=0\\
\frac{1}{V_f}\tilde \nabla&^\alpha_{\  A}(V_f\del_{\alpha}\phi)\\
&= \frac{e^\phi}{24}\tilde F^i_{\mu\nu}\tilde F^{i\mu\nu}+\frac{e^{-\phi}}{4}+\frac{1}{2V_f}\tilde\square_A V_f+\frac{\del\ln V_f}{3\del\phi}R\label{deltaphi}\\
T^{YM}_{\alpha\beta}&=\frac{1}{2}\left(\tilde g^{\mu\nu}\tilde F^i_{\alpha \mu}\tilde F^i_{\beta \nu}-\frac{1}{4}\tilde g_{\alpha\beta}\tilde F^i_{\mu\nu} \tilde F^{i\mu\nu}\right)\\
T^\Phi_{\alpha\beta}&= 3\left(\frac{1}{4}\tilde g_{\alpha\beta}e^{-\phi}-\frac{1}{2}\tilde g_{\alpha\beta}\del_\mu \phi\del^\mu\phi+\del_\alpha \phi\del_\beta \phi\right.\\
&\quad\quad\left.+\frac{1}{2} (\tilde g_{\alpha\beta}(\del_\mu\phi) \del^\mu\ln V_f-(\del_{(\alpha}\phi)\del_{\beta)}\ln V_f)\right).
\end{align*}
which for $\phi$ constant still leads  to an unphysical constraint on the YM field, now on the size of $||\tilde F||^2$. 

In our case, however, because the extra dimensions are noncommutative,  we propose that we cannot compute the equations of motion in the same way as we would for classical fields. Even though our fields were constrained by the requirement of a bimodule inverse to not depend on the fuzzy sphere coordinates, they originate on the product and need to be varied as such. Variational calculus on in noncommutative geometry is not understood, requiring first a better understanding of noncommutative jet bundles.  

\section{Lorentzian quantum gravity on a fuzzy sphere fibre} \label{secqg}

An alternative work-around, which we now explore, is to dispense with the equations of motion attributable to variation in the fibre direction and  instead quantise the metric modes in the fibre direction. We focus on  the  $R_h$ term in the action that depends only on the fuzzy sphere and which should be the relevant Einstein-Hilbert action for the fibre. Our approach to the other terms, in first approximation, will be (i) we ignore them for the quantum gravity on the fuzzy sphere on the grounds that, while they could be viewed as sources coming from the background spacetime fields, these should be insignificant compared to the Planck scale relevant to quantum gravity (ii) after quantising gravity on the fibre, we replace $h_{ij}$ in these other terms by (some version of) its expectation value, resulting in an effective theory of $\tilde A,\tilde g$ fields on $M$ as scales well above the Planck scale. In this approximation we can hope that the fibre quantum gravity, whatever it is, should be rotationally invariant and hence
\[ \<h_{ij}\>=h(x)\delta_{ij}\]
for some function $h(x)$ (where we suppose that the theory is quantised independently at each spacetime point). The issue here is that the relevant quantum gravity theory will have divergences and need to be regularised and renormalised. We need to know that we can make sense of this in a way that preserves the symmetry. We will then have significant freedom in the renormalisation process and we propose that the quantum gravity theories at different spacetime points in $M$ would naturally be regularised and renormalised uniformly, so as to have a constant value of $h$ independently of spacetime. This scenario would then complete the derivation of the Kaluza-Klein picture as coming out of quantum gravity on the fibre and sidestepping the problem with the equations of motion. 

The first problem we have here is that although quantum gravity on the fuzzy sphere with the action $R_h$ was already studied in \cite{LirMa} and in more depth in \cite{Ma:qg}, this was as Euclidean quantum gravity without an $\imath$ in the action. Because, now, $R_h$ enters as part of the total Lorentzian Einstein-Hilbert action, the relevant effective theory inherits a Lorentzian $\imath$ in the action, which makes it significantly different. The other difference from previous work is that we shall retain an open mind about that should be the measure of integration on the space of metrics that we quantise over. We still use the idea in \cite{LirMa,Ma:qg} that as long as we are interested in observables that depend only on the eigenvalues of the metric as a positive matrix, it suffices to limit attention to diagonal metrics $h_{ij}={\rm diag}(\lambda_1,\lambda_2,\lambda_3)$. The partition function then looks like
\begin{align*} Z&=\int_\eps^L \extd^3\mu\  e^{{\imath\over 2 G}(\lambda_1^2+\lambda_2^2+\lambda_3^2-2(\lambda_1\lambda_2+\lambda_2\lambda_3+\lambda_3\lambda_1))},\end{align*}
where we cut off at both large and small field strengths and we take the $V_f=\int_{A_f}1=\det(h)=\lambda_1\lambda_2\lambda_3$. We consider 3 choices of measure
\[ \extd^3\mu=\extd\lambda_1\extd\lambda_2
\lambda_3 \begin{cases}1 & \rm (naive)\\ {1\over\lambda_1\lambda_2\lambda_3} & \rm (Liouville) \\  {\small {|(\lambda_1-\lambda_2)(\lambda_2-\lambda_3)(\lambda_3-\lambda_1)|\over\lambda_1^2\lambda_2^2\lambda_3^2}} & \rm (geometric),\end{cases}\]
where the Liouville case corresponds to $\extd\phi_i$ for  $\phi_i=\ln \lambda_i$. The `geometric' measure in \cite{LirMa,Ma:qg} is the Riemannian measure on the space of real symmetric matrices as a symmetric space. 

\subsection{Lorentzian QG with the naive measure} 
\begin{figure}[h]\[ \includegraphics[scale=0.65]{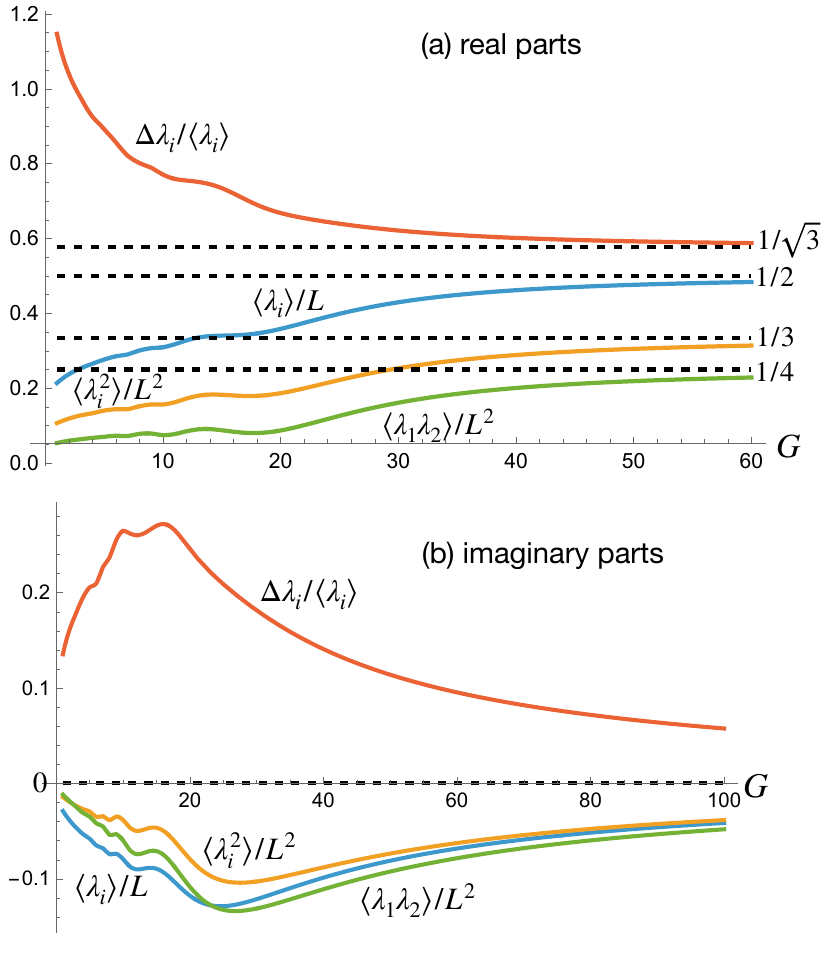} \] 
\caption{Lorentzian fuzzy sphere quantum gravity expectations with Liouville measure and cutoffs $L$ and  $\eps=10^{-5}$, as a function of coupling constant $G$. The dashed lines are the relevant $G\to\infty$ limits. \label{fig1} } 
\end{figure}

In this case two of the integrals, say $\lambda_1,\lambda_2$, can be done analytically. The final integral, $\lambda_3$ will then be doable provided ${\rm Im}(G)<0$. We take the latter as a prescription in which we afterwards use the result for real $G$.  For the partition function the result is 
\begin{align*} &z(\lambda_3;\eps,L,G)\\
&=\sqrt{\pi G^3\over 2}{e^{\imath\pi\over 4}\over\lambda_3} \Big(e^{-\frac{2 i \lambda_3 \eps}{G}}  \big(\text{erfi}(\frac{e^{\imath\pi\over 4} (\eps-L+\lambda_3)}{\sqrt{2G}})-\text{erfi}(\frac{e^{\imath\pi\over 4}\lambda_3}{\sqrt{2G}})\big)\\
&+ e^{-\frac{2 i \lambda_3 L}{G}} \big(\text{erfi}(\frac{e^{\imath\pi\over 4} (L-\eps+\lambda_3)}{\sqrt{2G}})-\text{erfi}(\frac{e^{\imath\pi\over 4} \lambda_3}{\sqrt{2G}})\big)\Big).\end{align*}
There are similar (but more complicated) expressions with $\lambda_1,\lambda_1\lambda_2,\lambda_1^2$  inserted in the integrand for $\<\lambda_1\>,\<\lambda_1\lambda_2\>,\<\lambda_1^2\>$ etc. (and the same for  $\lambda_1$ swapped with $\lambda_2$  by symmetry). The final integration for the partition function,
\[ Z=\int_\eps^L\extd\lambda_3 z(\lambda_3;\eps,L,G)\]
 is then done numerically, similarly with $\lambda_3$ inserted in this integral for $\<\lambda_3\>$, etc. This allows us to compute the total partition functions $Z$ and expectations $\<\lambda_1\>=\<\lambda_2\>$ and $\<\lambda_3\>$ as functions of $G$ with $\eps,L$ fixed,  as shown in Figure~\ref{fig1}. $Z$ itself is log divergent in $L$ for a fixed $G$ and also divergent in $G$ for a fixed $L$.
 
In fact  $Z$ and  expectation values involving a positive power of $\lambda_3$ are nonsingular at $\eps=0$, so  one could just set this, but for expectations involving only $\lambda_1,\lambda_2$  the numerical integrand becomes too highly oscillatory for smaller $G$, namely when $g<0.095$ for onset of the first such instability, where 
\[ g:=G/L^2\]
in this section (where the IR cutoff dominates). This instability becomes increasingly severe as $\eps\to 0$. (These observations are for $\eps=10^{-10}$ and numerical integration with MATHEMATICA at MaxRecursion 100 and machine precision.) In the plots, we used $\eps=10^{-5}$ for which the results of the numerical integration appear accurate to within 1\%  of numerical noise throughput the range of $G$.  The key check is that $\<\lambda_1\>, \<\lambda_1\lambda_2\>, \<\lambda_1^2\>$  are nevertheless not visibly different from $\<\lambda_3\>, \<\lambda_1\lambda_3\>, \<\lambda_3^2\>$ respectively as plotted in  Figure~\ref{fig1}. This confirms that our regulatory scheme preserves the symmetry between the $\lambda_i$. Our plots are for $L=10$ but the same shape applies for any scale as $\<\lambda_i\>/L$ and $\<\lambda_i\lambda_j\>/L^2$ depend only on $g$.  
 
 The $G\to\infty$ `deep quantum gravity' limit values are shown in the figure as dashed lines and given by real quantities
\[ \<\lambda_i\>\sim {L\over 2},\quad \<\lambda_i\lambda_j\>\sim  \begin{cases}L^2\over 4 & i\ne j\\ L^2\over 3  & i= j\end{cases},\quad {\Delta\lambda_i\over\<\lambda_i\>}={1\over \sqrt{3}},\]
which just correspond to omitting the action entirely, and we see in the figure  that we approach these for 
\[ g>1. \]
Here $\Delta\lambda_i/\<\lambda_i\>=\sqrt{\<\lambda_i^2\>/ \<\lambda_i\>^2-1}$ is complex for finite $G$ (this is typical in a Lorentzian theory)   but approaches the limiting real value stated at large $G$.  We did not see the phase transition observed in the Euclidean theory in \cite{Ma:qg} (for the geometric measure and in other conventions where we set $G=1$). Note that the large $L$ expansions there correspond to weak $G\to 0$ expansions here, or more precisely expansions of $\<\lambda_i\>/L$ and $\<\lambda_i\lambda_j\>/L^2$ as functions of $g$ for small $g$. For example
\begin{equation}\label{z3appx} \<\lambda_3\>/L\approx -{1.28\over \ln(g)}- {0.6\imath\over  \ln(g)^{2}}\end{equation}
is a good approximation on a log-linear plot for small $g$ down to machine limitations of $10^{-17}$. We see that, while complex, this expectation value tends to real as $g\to 0$ or $L\to \infty$. We will also be interested in the value of $Z$ itself for small $g$ and a similar analysis gives 
\begin{equation}\label{Znappx} |Z|\approx 10.77\sqrt{-\ln(g)}G^{3\over 2}\end{equation}
as a good fit  down to machine limitations around $g=10^{-14}$. The phase appears to tend to approximately $e^{-\imath\pi/4}$ but will not be needed. Both  small $g$ expansions were done for $\eps=0$ as UV regulation was not needed for these quantities.

\subsection{Lorentzian QG with Liouville measure}

Taking the Liouville measure, the partition function is
\begin{align*} Z&=\int_\eps^\infty \frac{\extd\lambda_1\extd\lambda_2
\extd\lambda_3}{\lambda_1\lambda_2\lambda_3}  e^{{\imath\over 2 G}(\lambda_1^2+\lambda_2^2+\lambda_3^2-2(\lambda_1\lambda_2+\lambda_2\lambda_3+\lambda_3\lambda_1))}.\end{align*}
Here we set $L=\infty$ as $Z$ is non-singluar at $L=\infty$ and this is also true for the expectation values for $\lambda_i,\lambda_i\lambda_j$. The partition function $Z$ is, however, log divergent as $\eps\to 0$. By contrast, one can show that $\<\lambda_i\>=\<\lambda_i^2\>=\<\lambda_i\lambda_j\>\to 0$ as $\eps\to 0$, while ratios such as $\frac{\Delta\lambda_i}{\<\lambda_i\>}$ can be expected to be log-divergent.  

For $\<\lambda_1\>$, we first do the $\lambda_1$ integral provided ${\rm Im}(G)<0$, which as for the naive measure case, we take as a prescription and then use the result for real $G$. This gives
\begin{align*} z1&(\lambda_2,\lambda_3; \eps, G)\\
&=\sqrt{\pi G\over 2}\frac{e^{-{\imath\pi\over 4}}}{\lambda_2\lambda_3} e^{-\frac{2 i \lambda_2\lambda_3}{G}} \text{erfi}\left(\frac{e^{\imath\pi\over 4} (\lambda_2+\lambda_3-\eps)}{\sqrt{2G}}\right),\end{align*}
which we then integrate numerically for $\int_\eps^\infty \extd \lambda_2\extd \lambda_3$. There are similar analytic expressions for $z11, z12$ for the calculations of $\<\lambda_1^2\>, \<\lambda_1\lambda_2\>,$ respectively. The denominator $Z(\eps,G)$, however, has to be done numerically. Similarly doing $\<\lambda_i\>,\<\lambda_i^2\>$ etc fully numerically produced non-convergence errors in Mathematica resulting in a significant degree of numerical `noise' but in broad agreement with the more precise hybrid method described.

\begin{figure}[h]\[ \includegraphics[scale=0.6]{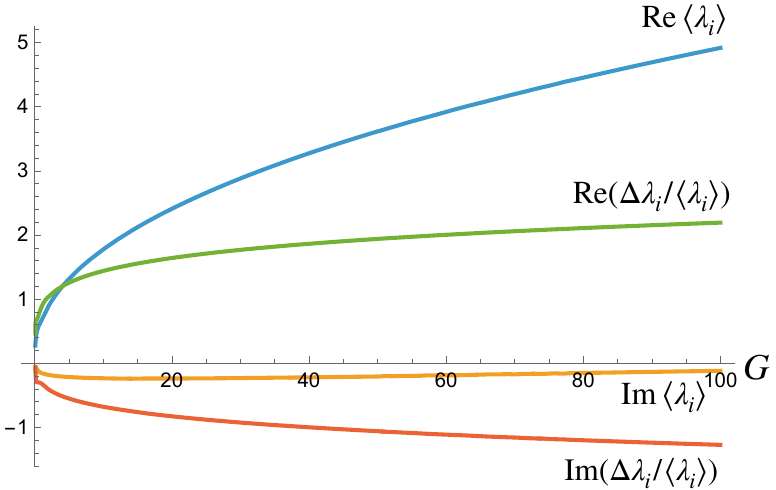} \] 
\caption{Lorentzian fuzzy sphere quantum gravity expectations with Liouville measure and cutoff  $\eps=0.1$, as a function of coupling constant $G$.  \label{fig2}}
\end{figure}

Results are shown in Figure~\ref{fig2} for $\eps=0.1$ but the same shape applies for all $\eps$ as $\<\lambda_i\>/\eps$ and $\<\lambda_i^2\>/\eps^2$ etc. depend only on 
\[ g:= G/\eps^2\]
in this section (where the  UV cutoff dominates).  Here, $Z(\eps,G)=Z(1,G/\eps^2)$ by rescaling the integration variables, and similarly for the expectations with a power of $\eps$. We omitted showing $\<\lambda_i\lambda_j\>$ but their plots are similar to that of $\<\lambda_i\>$.

The expectations are again complex with $\<\lambda_1\>$ oscillatory on a scale that depends on $g$. For example, there is a value of $g$ where $\<\lambda_i\>$ {\em is} real, off the scale in Figure~\ref{fig2} at $G=174.6$ but one can see the  approach. We will also need the value of $Z$ itself for small $g$. Here, numerical methods were not convergent for $g<0.1$, but we can instead use analytic methods as follows. First, we change variables from $\lambda_i$ to the Liouville field $\phi_i=\ln(\lambda_i)$. Then 
\begin{align*}Z(&\eps,G)=Z(1,g)\\
&=\int_0^\infty\extd^3\phi\ e^{{\imath\over 2g}(e^{2\phi_1}+e^{2\phi_2}+e^{2\phi_3}-2(e^{\phi_1+\phi_2}+e^{\phi_2+\phi_3}+e^{\phi_3+\phi_1}))}\\
&\approx e^{-{3\imath\over 2 g}}\int_0^\infty\extd^3\phi\ e^{-{\imath\over g}(\phi_1+\phi_2+\phi_2+\phi_1\phi_2+\phi_2\phi_3+\phi_3\phi_1)} \end{align*}
where we expand for small $\phi_i$ to quadratic order. The critical point for the action here is at $\phi_i=-\infty$ and moreover the Hessian has vanishing determinant at any symmetric value of the $\phi_i$, hence one cannot use the stationary phase method in the usual way. Instead, the integral will be dominated by the boundary at $\phi_i=0$. This results in
\begin{equation}\label{Zlapprox} Z\approx e^{{\imath\over 2}(\pi- {3 \over g})}g^3 \end{equation}
on numerical evaluation, as a good approximation for $g<< 0.1$ down to machine precision. We obtain the same conclusion if we expand the original action to quartic degree, justifying the approximation. We can use the same technique for the leading form of $\<\lambda_3\>=\<e^{\phi_3}\>$, say, for small $g$. The $\phi_3$ integral can be done analytically assuming ${\rm Im}(g)<0$ as a prescription, and the remainder then done numerically using the result with $g$ taken real. Dividing by $Z$ from (\ref{Zlapprox}) for the expectation value, we find
\begin{equation}\label{Zl1approx} \<\lambda_3\>\approx \eps e^{\imath{3 \over 2g}}\end{equation}
for $g<<0.1$, which is to say, small $G$ for a fixed $\eps$.

\subsection{Lorentzian QG with geometric measure}

\begin{figure}[h]\[ \includegraphics[scale=0.6]{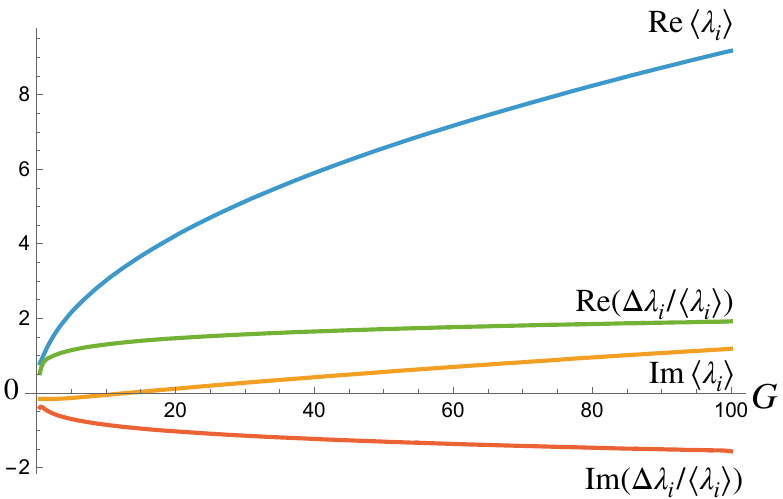} \] 
\caption{Lorentzian fuzzy sphere quantum gravity expectations with geometric measure and cutoff  $\eps=0.1$, as a function of coupling constant $G$.   \label{fig3}}
\end{figure}

For the geometric measure, we use direct numerical calculation of the partition function and for $\<\lambda_1\>$, which turns out to be much more stable if we work with inverted coordinates. We are again able to integrate to $L=\infty$ with the divergences at zero regulated by $\eps>0$, by the same arguments as in the Liouville measure case. Thus, the partition function is
\begin{align*} Z&=\int_\eps^\infty \extd\lambda_1\extd\lambda_2
\extd\lambda_3{|(\lambda_1-\lambda_2)(\lambda_2-\lambda_3)(\lambda_3-\lambda_1)|\over\lambda_1^2\lambda_2^2\lambda_3^2}\\
&\qquad e^{{\imath\over 2 G}(\lambda_1^2+\lambda_2^2+\lambda_3^2-2(\lambda_1\lambda_2+\lambda_2\lambda_3+\lambda_3\lambda_1))}\\
&=\int_0^{1/\eps} \extd l_1\extd l_2
\extd l_3{|(l_1-l_2)(l_2-l_3)(l_3-l_1)|\over l_1^2 l_2^2 l_3^2}\\
&\qquad e^{{\imath\over 2 G}\left(\frac{1}{l_1^2}+\frac{1}{l_2^2}+\frac{1}{l_3^2} -2\left(\frac{1}{l_1 l_2}+\frac{1}{l_1 l_3}+\frac{1}{l_2 l_3}\right)\right)}\end{align*}
where we set $l_i=1/\lambda_i$, and similarly with a $1/l_i$ factor when computing the numerator of $\<\lambda_i\>$. The results for $\<\lambda_i\>$ are shown in Figure~\ref{fig3} against $G$ and for $\eps=0.1$, but as for the Liouville measure case, only the ratio $g:=G/\eps^2$ is relevant since  $Z(\eps,G)=Z(1, G/\eps^2)$, and similarly for other expectations with a power of $\eps$.  In passing we note that the real part of the numerator of $\<\lambda_i\>$ can be modelled as $7.7 G/\eps$ to a good approximation as $\eps\to 0$ or $G\to \infty$ (the imaginary part was less clear from the data available). 

We now focus on the numerator of the ratio $\<\lambda_1^2\>$. Following the strategy of the Liouville case , we are able to do the integration $\extd \lambda_1$ analytically for the original partition function now with an extra $\lambda_1^2$, provided ${\rm Im}(G)<0$, which, as before, we take as a prescription and then use the result for real $G$. The only subtlety is that we integrate with $(\lambda_1-\lambda_2)(\lambda_1-\lambda_3)$ in the integrand, without the absolute value. This factor would anyway be positive for $\lambda_1<m_3$ or $\lambda_1>m_2$ and negative otherwise, where $m_2:=\max(\lambda_2,\lambda_3)$ and $m_3:=\min(\lambda_2,\lambda_3)$. Hence we integrate $\lambda_1\in (\eps,\infty)$ and subtract twice the integral for $\lambda_1\in (m_3,m_2)$. There integrals can be done analytically to give 
\begin{widetext}
\begin{align*} f(\lambda_2,\lambda_3, m_2,m_3)&=\frac{(m_2-m_3)G}{ \lambda_2^2\lambda_3^2} e^{-\frac{i (\eps (\lambda_2 +\lambda_3 )+2 \lambda_2  \lambda_3 )}{G}} \Big( \imath \eps e^{\frac{i (\eps^2+(\lambda_2 +\lambda_3 )^2)}{2 G}}+2 \imath  m_2   e^{\frac{i (2 \eps (\lambda_2 +\lambda_3 )+m_3 ^2)}{2 G}}-2 \imath  m_3  e^{\frac{i \left(2 \eps (\lambda_2 +\lambda_3 )+m_2  ^2\right)}{2 G}}\\
&\quad +e^{\imath\pi\over 4}\sqrt{\pi \over 2 G} (G-i \lambda_2  \lambda_3 ) e^{\frac{i \eps (\lambda_2 +\lambda_3 )}{G}} \big(\imath + \text{erfi}(\frac{e^{\imath\pi\over 4} (-\eps+\lambda_2 +\lambda_3 )}{\sqrt{2G}})+2 \text{erfi}(\frac{e^{\imath \pi\over 4} m_3 }{\sqrt{2 G}})-2 \text{erfi}(\frac{ e^{\imath \pi\over 4} m_2  }{\sqrt{2 G}})\big)\Big).\end{align*}
\end{widetext}
The  integrand also has a factor  $|\lambda_2-\lambda_3|/(\lambda_2^2\lambda_3^2)=(m_2-m_3)/(\lambda_2^2\lambda_3^2)$ as shown. We then integrate this  numerically over $\lambda_2,\lambda_3$ for the two regions $\lambda_2\in (\eps,\infty)$, $\lambda_3\in (\lambda_2,\infty)$ where we use $m_2=\lambda_3, m_3=\lambda_2$, and $\lambda_3\in (\eps,\infty)$, $\lambda_2\in (\lambda_3,\infty)$, where we use $m_2=\lambda_2, m_3=\lambda_3$. These two regions contribute equally due to the symmetry between $\lambda_2,\lambda_3$, generating a factor 2 times the second case. The numerator of $\<\lambda_1^2\>$ is therefore the integral 
\[ 2\int_\eps^\infty\extd\lambda_3 \int_{\lambda_3}^\infty\extd\lambda_2 f(\lambda_2,\lambda_3,\lambda_2,\lambda_3)\]
which we again do numerically. The end result for the relative uncertainty is also plotted for   $\eps=0.1$ in Figure~\ref{fig3} (other values have the same shape due to the scaling). 

We also find that 
\[ \<\lambda_1^2\>\approx {G^2\over\eps^2 Z}\]
 to a  good approximation as  $\eps
\to 0$ or $G\to \infty$. As for the Liouville measure case, we did not find real expectations for either large or small $G$ (to the limited extent visible numerically) but there is a critical value $g=1315$ where $\<\lambda_i\>$ {\em is} real, visible at $G=13.15$ in Figure~\ref{fig3}.

\section{Matching to physical values} \label{secmatch}

In the preceding section, we studied quantum gravity in the fuzzy sphere by the same methods as in \cite{LirMa} but now with a Lorentzian action. Moreover, because the measure on the space of metrics is additional data for which we do not have a definitive theory, we did this for three natural choices. The first measure was in line with the discrete quantum gravity models such as \cite{Ma:squ,BliMa} where we regard the metric `square length' values as the variables and we obtained results similar to such models. The model had an essential IR cut off at large metric values $L$. As a result, these expectation values in the bare theory are related to $L$ and there was a  $G\to\infty$ limit with identified asymptotes and a limiting value $\Delta \lambda_i/\<\lambda_i\>= 1/\sqrt{3}$ of the relative uncertainty. These limits are the same as for the Euclidean version of the theory in \cite{Ma:qg} and can also be obtained for the other two measures {\em provided} we cut these off at fixed $L$ and only then send $G\to \infty$. By rescaling the fields, the large $G$ theory is equivalent to $L\to 0$, but we did not find signs of a phase transition below a critical value as for the Euclidean theory in \cite{Ma:qg}. For the most important observables of interest, we could set $\eps=0$ for the UV cutoff of small metric field strengths. 

By contrast the Liouville and geometric measure cases did not need such an IR regulator and we set $L=\infty$. The two theories scale the same way on change of the remaining UV cutoff $\eps$ by rescaling the field variables, with result that $\eps\to 0$ is equivalent to $G\to \infty$. These measures were  motivated respectively from $\phi_i=\ln(\lambda_i)$ as the Liouville field variable or from the space of metrics as a Riemannian manifold. These two measures gave remarkably similar results, this time without clear  $G\to \infty$ asymptotes.

Fortunately, the main things we want to take away held up in all three models: (i) for a fixed the integration scheme (which could have broken the symmetry between the $\lambda_i$) our results for $\<\lambda_i\>$ were nevertheless independent of $i$, corresponding to a sphere for the expected value for the metric. (ii) Due to divergences, at one end or the other, the theory has to be regularised and in principle renormalised. The idea is that in any such renormalisation scheme there will be freedom in which the observed value at some length scale is matched to physical values, after which the the renormalisation scheme determines the value at other scales. In our case, the relevant scale would be that of the Standard Model and the physical value would be determined by matching to the relevant Newton and Yang-Mills coupling constants. In this case, it would be natural to do this in the same way at all points of spacetime. Hence, $\<h_{ij}\>$ would emerge naturally as constant on spacetime due to the  inherent freedom in the renormalisation process.  

We now look at this is in more detail, focussing on the naive and Liouville measures. The first wrinkle is that the relevant expectation values are not necessarily real. Our approach to this will be to take the absolute value for the purposes of matching, and look where possible for approximate reality in  the relevant limit. The base scenario is then to take the effective metric in the Yang-Mills part of (\ref{action}) to be $h_{ij}:=h\delta_{ij}$ with  $h=|\<\lambda_i\>|$. We also use this classical value for all other terms in the action that involve spacetime fields, so $V_f=h^3$ for the determinant of this metric. We look at the total theory with coupling constant $G_0$ say (so $\imath/G_0$ times the action (\ref{action}) for the exponent in the functional integral). Then we match the gravitational part of the action by 
\begin{equation}\label{Gval} G_0=8\pi V_f G_N= 8 \pi V_f l_p^2\end{equation}
where $G_N$ is Newton's constant and we set $c=\hbar=1$. We also allowed for the fact that $\tilde{R}$ in our conventions is -1/2 of the usual value, ignoring the minus since in the Lorentzian theory it is equivalent via complex conjugation. Next, as in \cite{LiuMa2}, we look at the Yang-Mills term and need 
\begin{equation}\label{hval}  h:=\frac{16\pi l_p^2}{g^2_{YM}}\end{equation}
in order to have the right ratio between the GR and Yang-Mills sector. As shown there, to match the electroweak theory, for example, we need $\sqrt{h}=11 l_p$.  Finally, we look at the fuzzy sphere action $R_h$ term in (\ref{action}) which in terms of $\lambda_i$ as in \cite{LirMa} and cancelling $V_f$ gives the action we have used in this section with
\begin{equation}\label{GG0}  G={2 G_0 \over  V_M }=16\pi V_f {l_p^2\over l_u^4} \end{equation}
where $V_M=\int_M\extd^4x \sqrt{-|\tilde g|}=l_u^4$
say for a cosmological scale $l_u$. If we take the actual size of the Universe then this would be $5.4\times 10^{61}l_p$ in Planck scale units, but realistically one can take any $l_u$ much bigger than the length scales of interest.

\subsection{Matching for naive measure with $h=|\<\lambda_i\>|$}\label{secmatchn1}

We first look at the naive measure where we can set $\eps=0$ and cut off at  $L$ for a maximum square-length scale in the fuzzy sphere quantum gravity. Using the rescaling of fields,  we need 
\[ |\<\lambda_i\>_{L, G}|= L|\<\lambda_i\>_{1, {G\over L^2}}| = h\]
which we can solve for $G$, presumably in the very weak $G$ regime in Figure~\ref{fig1} as approximated in (\ref{z3appx}).  If we assume this then 
\begin{equation}\label{naiveGL}   G\approx L^2 e^{ -1.28 \frac{ L}{h}} \end{equation}
for  $L>> h$. However, if we take $V_f=h^3$ as in the base scenario then  we also need
\[ G= 16 \pi {l_p^2 h^3\over l_u^4}\]
and equating these gives
\[ e^{-1.28 {L\over h}}({L\over h})^2 =16 \pi   {h\over l_p^2}  \big({l_p\over l_u}\big)^4\]
which for $l_u/l_p=5.4 \times 10^{61}$ and $h=121 l_p^2$ would give $L/h$ as 447 or
\begin{equation}\label{Lnaivea} \sqrt{L}\approx 21 \sqrt{h}\end{equation}
or about $230 l_p$, which is a respectable upper cutoff for for quantum gravity on the fibre, i.e. well above the Planck scale.  The value if $L/h$ here is relatively insensitive to the values of $l_u,h$ due to the exponential.  

Note that we should also consider the renormalised theory, and the approach as in \cite{Ma:qg} is to match a physically observed quantity. In our case the physically observed quantity is $h$ and choosing $G(L)$ so that $|\<\lambda_i\>_{L,G(L)}|$ lands on it is exactly what we did as solved by (\ref{naiveGL}). However, we also wanted the running coupling $G(L)$ to land on the observed $G_N$, which in our units fixed $L$ as stated rather than being able to remove the regulator entirely. In principle, we could declare that $G_N$ also depends on the scale $L$ but this is less clear given that we work in units $c=\hbar=1$; it would need the effective Planck length to also run.

\subsection{Matching for naive model with $h=|\frac{\<\lambda_1\lambda_2\lambda_3\lambda_i\>}{\<\lambda_1\lambda_2\lambda_3\>}|$}\label{secmatchn2}

\begin{figure}[h]\[ \includegraphics[scale=0.9]{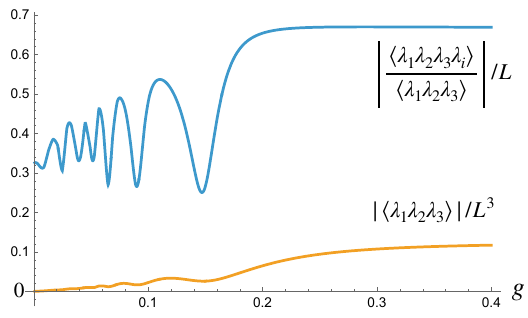} \] 
\caption{Further Lorentzian fuzzy sphere quantum gravity expectations with naive measure as a function of $g=G/L^2$.   \label{fig4}}
\end{figure}

We now look at a second scenario where we do not treat treat $V_f=\det(h)$ as classical but use $\<\lambda_1\lambda_2\lambda_3\>$ for the gravitational part of the action and $\<\lambda_1\lambda_2\lambda_3\lambda_i\>/\<\lambda_1\lambda_2\lambda_3\>$ for the Yang-Mills part of the action relative to the gravitational part.  Some plots for these expectations are shown in Figure~\ref{fig4}.  We see that
\[ \left |{\<\lambda_1\lambda_2\lambda_3\lambda_i\>\over \<\lambda_1\lambda_2\lambda_3\>}\right|\to {L\over 3}\]
as $G\to 0$. Hence with $h$ denoting the observed ratio (\ref{hval}), we need
\[ \sqrt{L}\approx 1.7 \sqrt{h}\]
or $20 l_p$ for $h$ to match the electroweak case. This is borderline as a cutoff for quantum gravity on the fibre, but would fit better for much larger $h$ as could apply for example for $SU(2)_f$ flavour symmetry (as discussed  in Section~\ref{seccon}). We now also need
\[ G= {16\pi l_p^2\over l_u^4}|\<\lambda_1\lambda_2\lambda_3\>|={16\pi l_p^2 L^3\over l_u^4}|\<\lambda_1\lambda_2\lambda_3\>_{1,{G\over L^2}}|\]
and we find numerically for small $g$ that
\[ {\<\lambda_1\lambda_2\lambda_3\>\over L^3}\approx - 0.035 g^{1.2}- \imath 0.25 g^{1.1}\]
appears to be a fair approximation to $g=G/L^2$ approaching machine limitations of $10^{-6}$. At small $g=G/L^2$, the imaginary part dominates and using the size of this requires 
\[  ({G\over L^2})^{0.1}=({l_u\over l_p})^4 {l_p^2\over 4\pi L}\]
hence $g<<1$ needs
\[ \sqrt{h} >0.16\  l_p ({l_u\over l_p})^2 \]
which is unreasonable. Even take a much larger $\sqrt{h}\approx 1.7\times 10^{16} l_p$ relevant to a possible flavour symmetry $SU(2)_f$, this translates to 
\[ l_u< 3.2\times 10^8 l_p\]
which is unreasonable for the standard model.  On the other hand, these particular calculations are very sensitive to the modelling of the small $g$ behaviour and our extrapolation to the much smaller $g$ needed, hence should be seen as only indicative.

We have a similar issue for large $g>>1$, where we find
\[ \<\lambda_1\lambda_2\lambda_3\>\to {L^3\over 8},\quad {\<\lambda_1\lambda_2\lambda_3\lambda_i\>\over \<\lambda_1\lambda_2\lambda_3\>}\to {2\over 3}L\]
so $\sqrt{L}=1.2 \sqrt{h}$ similar to before, while we require
\[ {G\over L^2}\approx 2 \pi {l_p^2\over l_u^4}L,\]
which contradicts $g>> 1$ unless 
\[ \sqrt{h}>> 0.32\ l_p({l_u\over l_p})^2\]
and hence leads to a similar conclusion as before. In short, this second scenario for $V_f$ does {\em not} appear to work well for physical values if
we use the naive measure and want to land on the observed coupling for gravity with a reasonable box size $l_u$.

These expectation values are complex, but $\<\lambda_1\lambda_2\lambda_3\lambda_i\>/\<\lambda_1\lambda_2\lambda_3\>$ tends to real for both large and small $g=G/L^2$, while  $\<\lambda_1\lambda_2\lambda_3\>$ tends to real for large $g$ and appears to tend to imaginary for small $g$. There are also repeated intermediate values of $g$ where the former is  real and ditto where the is exactly imaginary,  increasingly often as $g\to 0$.

\subsection{Matching for the Liouville measure}\label{secmatchl}

There is a parallel calculation for the other two measures, which have a main UV regulator $\eps$ and hence behave  differently from the preceding section. We check this for the Liouville case, but the geometric measure appears to be broadly similar and is not expected to reach a fundamentally different conclusion. If we follow the original base scenario  then using the small $G/\eps^2$ approximation (\ref{Zl1approx}) where $ |\<\lambda_i\>|\approx \eps$, we have
\[ h=\eps,\quad V_f=\eps^3,\quad {G\over\eps^2}=16\pi h {l_p^2\over l_u^4} \]
valid for 
\[ h< < {l_u^4\over 16\pi l_p^2 }.\]
In particular, this is valid for any $l_u$ much bigger than $\sqrt{h}$ and $4\sqrt{\pi} l_p$, i.e. certainly for any macroscopic scale. This therefore matches well to our requirements. 

The second scenario is to set 
\[ {|\<\lambda_1\lambda_2\lambda_3\lambda_i\>|\over |\<\lambda_1\lambda_2\lambda_3\>|}=|\<\lambda_i\>_{naive}|=h\]
which we have already covered in Section~\ref{secmatchn1}  (these expectations as we saw need IR regulation by $L$ and we can set $\eps=0$). We recall from there that $G/L^2$ from (\ref{naiveGL}) will be extremely small in practice. For $V_f$ however, we now need 
\[ V_f= |\<\lambda_1\lambda_2\lambda_3\>|={|Z_{naive}(L, G)|\over |Z_{Liouv}(\eps,G)|}\approx 10.77 {\eps^6 \over G^{{3\over 2}}}\sqrt{\ln(L^2/G)}.\]
using (\ref{Znappx}) for the numerator and (\ref{Zlapprox}) for the denominator.  The two requirements are therefore (\ref{naiveGL}) as before, and, given this, the $V_f$ condition (\ref{GG0}) becomes
\begin{align*} {e^{-{5\over 2}1.28 {L\over h}} \over \sqrt{1.28 {L\over h}}}({L\over h})^5=16\times 10.77 \pi  {\eps^6 l_p^2\over h^5 l_u^4},\end{align*}
where we eliminate $G/L^2$.  The value of $\eps$ is free except that for (\ref{Zlapprox}) we need $G<<\eps^2$, which is clear given the exponential in (\ref{naiveGL}). For example,  the electroweak case with minimum $\eps$ and maximum $l_u$ is 
\[  h=121 l_p^2,\quad \sqrt{\eps}=l_p,
\quad l_u=5.4 \times 10^{61}  l_p.,
\quad  \sqrt{L}\approx 13.8 \sqrt{h}\]
or about 152 $l_p$, which is a similar result to that in Section~\ref{secmatchn1} but now with Liouville measure and the second scenario where $V_f=|\<\lambda_1\lambda_2\lambda_3\>|$. The ratio $L/h$ here is again relatively insensitive to the parameters on the right hand side. 

For this second scenario, we already know that $\<\lambda_i\>_{naive}$ is very close to real for the values of $G/L^2$ in use, but $Z_{naive}, Z_{\rm Liouv}$ for small $G$ appear to be  oscillatory and repeatedly cross the real axis (so could be made real, or their ratio made real) for a discrete series of values of $L,\eps$ at a given $G$. 

\section{Concluding remarks}\label{seccon}

We have shown that $\<h_{ij}\>\propto \delta_{ij}$ in the Lorentzian fuzzy sphere quantum gravity with any of the three measures that we used and that we have enough freedom to take $h=|\lambda_i|$ to be constant as part of the control of divergences, as needed to be able to derive gravity plus Yang-Mills on spacetime from gravity on the total space with the fibre metric integrated out as quantum gravity on the fuzzy sphere fibre.  In the base scenario, we then use this value as an effective metric for rest of the action, which for example sets  $h=11 l_p$ if we want to match up with the electroweak Yang-Mills action in the case of the fuzzy sphere\cite{LiuMa2}. We showed in Section~\ref{secmatchn1} that this approach is viable and leads with the naive measure to a fuzzy sphere cutoff $L$ of around 200 $l_p$, giving  a reasonable dynamic range of integration of 1-200 $l_P$ for quantum gravity on the fibre. We found a very similar conclusion in Section~\ref{secmatchl} for the Liouville measure with the second scenario where we used the expectation values of $V_f=\lambda_1\lambda_2\lambda_3$ and $V_f\lambda_i$ in the action. The Liouville measure also produced a comfortable match for the base scenario with little restriction on $l_u$. Meanwhile, we found, in Section~\ref{secmatchn2}, that this second scenario can not match up for the naive measure, which is also important as it means that our approach is tightly constrained and falsifiable, working for some choices and not for others.  

The matching in Section~\ref{secmatch} depended on numerical methods which could only reach down to a certain value of $G$ due to machine limitations, whereas the actual value needed to match physical values turned out to require $G$ to be even smaller, requiring us to extrapolate. This could be looked at further with analytic methods. We also  had to contend with $\<\lambda_i\>$ etc., complex and saw at least in the first case (with the naive method) that this would be very close to real (and positive) for the relevant small $G$. Whilst we dealt with this by using absolute values, this question should be looked at further, either to further justify absolute values or to see if a very small imaginary component in the Yang-Mills coupling is of interest. 

Next, while the mechanism was illustrated for a fuzzy sphere fibre in \cite{LiuMa2}, where the symmetry of the fibre (quantum) geometry is $SU(2)$, our arguments leading to enforcing of the cylinder ansatz (and likely the rest of our proposed mechanism) also hold for other sufficiently noncommutative fibres. The main requirement is that the centre of the coordinate algebra  should be trivial and its calculus have a central basis. One should therefore look at other fibre quantum Riemannian geometries, e.g. ones with $SU(3)\times SU(2)\times U(1)$ symmetry, as another direction for further work. 

Moreover, while we have focussed on the geometric side, the QRG formalism also allows one to consider matter fields on the product. As in standard Kaluza-Klein theory with classical fibre, one can decompose such fields according to eigenvalues of the Laplacian for the fibre geometry, after which they appear as multiplets of fields on spacetime with different masses. This is an infinite tower in usual Kaluza-Klein theory with a classical compact fibre, but in our case if the fibre is a finite-dimensional noncommutative geometry then we would only have a finite multiplet. For example, using the reduced fuzzy sphere at $\lambda=1/2$ leads to a massless scalar field on the product appearing as an uncharged massless scalar and a massive $SU(2)$ Yang-Mills triplet (i.e. vector representation), see\cite{LiuMa2} for details. Likewise, a massless spinor field on the total space appears as two Yang-Mills  doublets and a novel Yang-Mills quadruplet, with induced masses in the distinctive ratio $1:5/3:7/3$, see \cite{LiuMa3}. 

Finally, we used the electroweak coupling constant for matching calculations, which requires $h$ and hence induced masses to approach Planckian. This is not necessarily without interest, for example Planckian mass Yang-Mills triplets have been used in connection with the seesaw mechanism for neutrino oscillations \cite{Gho}. A more novel possibility\cite{LiuMa2}, however, is that the Yang-Mills field corresponds to an as yet unobserved $SU(2)$ gauge symmetry.  This could be taken  to be ultraweak so that $\sqrt{h}$ falls in the TeV scale relevant to the Standard Model. Then the  induced masses of the lower non-zero spin  multiplets are at the Standard Model scale and we could potentially arrive at an explanation of some particle masses, for a suitable fibre quantum geometry.  The use of a noncommutative fibre would no longer be justified by quantum gravity effects, but could still be of interest in its own right. In the fuzzy sphere case, the associated $SU(2)$ gauge field would now have Yang-Mills coupling constant corresponding to an ultra-weak force, namely of order $10^{-32}$ in terms of dimensionless fine structure constant $\alpha$, see\cite{LiuMa2}. This is comparable to the self-gravity of the masses being created by the mechanism. 

Moreover, an area of the standard model where a new symmetry, possibly gauged, is likely needed would be to understand the `generations problem' where lepton families  are repeated as three generations. Notably, \cite{BerHer} propose a  `flavourspin' symmetry $SU(2)_f$ that mixes the three generations as spin 1 `vector' multiplets. This is a Minimal Flavour Violation (MVF) ansatz model and is not an exact symmetry, it would only be a symmetry at GUT scales (well above the fermion masses). The various Yukawa couplings in this context are  $M_3(\R)$-valued, i.e.,  $Y_{ij}$ with two generation indices, and enter into the Lagrangian as terms of the form
\[   \CL_{\rm Yuk}=-\overline{Q_L}Y_u U_R\cdot \tilde H- \overline{Q_L}Y_d D_R \cdot H - \overline{L_L}Y_l E_R\cdot H + h.c. \]
The various fermion fields here are grouped in threes due to a generation index $i$. Thus, $L_L$ denotes the left-handed leptons, $Q_L$ left-handed quarks, both doublets under the weak $SU(2)$, while $U_R,D_R,E_R,N_R$ are right-handed up, down, electron and neutrino fields respectively, which are singlets under the weak $SU(2)$, but all these fields are vectors under $SU(2)_f$. For example, $U_R$ contains right handed up, charm and top quarks as the triplet.  The idea, going back to  Cabbibo\cite{Cab} is to upgrade the Yukawa couplings to dynamical fields $Y_u,Y_d,Y_l, Y_\nu$ and such $Y_{ij}$ fields can indeed arise from a real scalar field $\phi$ on the product, which for the reduced fuzzy sphere at $\lambda=1$ can be decomposed into `flavourspin' $l=0,1,2$ components. In this case, the analysis in \cite{LiuMa2} would give the different spin components of each type of Yukawa field masses in the ratios $0:1:\sqrt{3}$ according to the values of $\sqrt{l(l+1)}$ if the real scalar field $\phi$ on the product is massless. Such model building is another direction for further work. At the moment this analysis is about proof of concept as we do not obtain matter fields and their masses matching the relevant sector of the Standard Model, but the ideas could be explored further.  A related direction for further work would be to understand the emergence of the Higgs field as claimed in  Connes' noncommutative geometry approach\cite{ChaCon}.  One could also consider quantum gravity plus matter on the fibre, i.e. quantise the fibre part of the matter fields, as a further related direction.

 \end{document}